# The toughness, elasticity, color, and consistency of cooperation
## Comment on "Reputation and reciprocity" by Xia *et al.*

**Petter Holme**
Department of Computer Science, Aalto University, Espoo, Finland
Center for Computational Social Science, Kobe University, Kobe, Japan
E-mail: petter.holme@aalto.fi

*This is a comment on a recent review article about reputation and reciprocity as mechanisms promoting cooperation. I also discuss the necessary changes for the currently game-theory-based cooperation studies to become a complete theory of cooperation in our contemporary society.*

The founder of the study of social interaction, Georg Simmel, once noted [1] that

> there exists an immeasurable number of less conspicuous forms of relationship and kinds of interaction. Taken singly, they may appear negligible. But since in actuality they are inserted into the comprehensive and, as it were, official social formations, they alone produce society as we know it. … Here are the interactions among the atoms of society. They account for all the toughness and elasticity, all the color and consistency of social life, that is so striking and yet so mysterious.

Cooperation is one of many classes of social interactions, all entwined into the social life Simmel describes. Clearly, the raw market forces and rational agents of neoclassical economics cannot account for all the toughness, elasticity, color, and consistency—we have to search for an understanding of cooperation in less conspicuous kinds of interaction. One step in this direction is to study reciprocity—the action of returning favors, directly or indirectly. This is a mechanism with a counterpart in the evolutionary theory of kin selection and, thus, hypothetically, a pre-historic trait. Another key mechanism behind cooperation is the social dynamics of reputation—how people collectively associate qualities with individuals that, subsequently, affect the individuals' participation in social processes.

The importance of studying human cooperation can hardly be overstated. The big issues facing humanity—global warming being an excellent example [2]—almost per definition require planet-wide cooperation and long-time perspectives. In their review [3], Xia *et al.* give a résumé of the research about how these mechanisms are manifested in experiments and how they drive cooperation in game-theoretic models. There is a vast range of plausible modifications [4,5] to the sub-mechanisms behind reciprocal action and reputation formation. Over the last decades, we learn, the field has proceeded by scanning through such scenarios. We now know much more about [3], for example: How the social network structure, and many other structural constraints on the interaction, affect these simple models of human interaction. The innate ability to judge fairness, and thus establish trust (through experiments with babies!). And, perhaps most prominently, the effects of additional behavioral information in decision-making.

From a decade-long perspective, the field described by Xia *et al.* [3] has some obvious challenges. Indeed, I will argue that with the current direction, resting heavily on game theory and, in extension, a world model populated by utility maximizers, we will never reach a complete theory of cooperation in modern humans (modern enough for economic success and genetic survival to be decoupled).

A first problem is that negative results, where cooperation is suppressed, are commonly unreported. So, we now have an extensive catalog of stylized situations where reputation and reciprocity, etc., promote cooperation, but the literature will not tell you whether the other situations are unstudied or destructive for cooperation. Second, the choice of what mechanisms to study is usually derived from earlier studies (experiments or theory) rather than observations or practical scenarios. This methodology of growing models outwards from the classics (the two-player prisoner's dilemma, etc.) is systematic but so glacial that one can wonder if we will ever reach beyond the current, rather extreme assumptions of a sociality shaped by economic forces.

The third and probably most serious issue is that we cannot easily take the results from experimen-tal or mathematical game theory as atomic facts and, by assembling these, predict real scenarios, as it would run into the frame problem [6]. I.e., in realistic situations, the number of behavioral factors that potentially could influence such a decision to cooperate explodes in number—religious faith [7], nourishment [8], anonymity [9], ethical considerations [10], personal identity [11], etc. [12] We can assume most of these potential factors and mechanisms would not even enter the mind of the decision maker, but there is no way of determining which ones to leave out without actually observing the scene unfold. Moreover, the before-mentioned publication bias exacerbates the problem as we don't have many negative mechanisms to choose from. Finally, almost nothing is known about how different atomic mechanisms (if it is even meaningful to talk about such) would interfere with each other.

The long-term issues aside, the study of human cooperation is thriving, and the research reviewed by Xia *et al.* [3] has many important discoveries ahead. I don't expect any revolution but rather a slow shift away from rational choice theory [4,13] and, by extension, game theory. Instead, I expect an increasing presence of methods from "evidence-based economics" [14], anthropology [15], and sociology [13]—more real-life observations, surveys, randomized control trials, and experiments that allows discovery of a wider range of mechanisms than today. Then, finally, we might understand the actions underpinning cooperation as entwined with the entire fabric of social interaction [15], "so striking and yet so mysterious."

## Acknowledgements

The author was supported by JSPS KAKENHI Grant Number JP 21H04595. Some ideas of this comment came out of discussions with Marko Jusup.